\documentclass[a4paper]{article}

\usepackage{dcolumn,booktabs}
\usepackage{pgf}
\usepackage{pgfplots}
\pgfplotsset{width=10cm,compat=1.9}
\usepackage{tipa}
\newcolumntype{d}[1]{D..{#1}}
\newcommand\mc[1]{\multicolumn{1}{c}{#1}} 

\usepackage{array}
\newcolumntype{P}[1]{>{\centering\arraybackslash}p{#1}}

\usepackage{INTERSPEECH2022}

\title{Deep Conditional Representation Learning for \\ Drum Sample Retrieval by Vocalisation}
\name{Alejandro Delgado$^1$, Charalampos Saitis$^1$, Emmanouil Benetos$^1$, Mark Sandler$^1$}
\address{
  $^1$Queen Mary University of London}
\email{a.delgadoluezas@qmul.ac.uk, c.sailtis@qmul.ac.uk, emmanouil.benetos@qmul.ac.uk, mark.sandler@qmul.ac.uk}

\begin{document}

\maketitle
\begin{abstract}
Imitating musical instruments with the human voice is an efficient way of communicating ideas between music producers, from sketching melody lines to clarifying desired sonorities. For this reason, there is an increasing interest in building applications that allow artists to efficiently pick target samples from big sound libraries just by imitating them vocally. In this study, we investigated the potential of conditional autoencoder models to learn informative features for Drum Sample Retrieval by Vocalisation (DSRV). We assessed the usefulness of their embeddings using four evaluation metrics, two of them relative to their acoustic properties and two of them relative to their perceptual properties via human listeners' similarity ratings. Results suggest that models conditioned on both sound-type labels (drum vs imitation) and drum-type labels (kick vs snare vs closed hi-hat vs opened hi-hat) learn the most informative embeddings for DSRV. We finally looked into individual differences in vocal imitation style via the Mantel test and found salient differences among participants, highlighting the importance of user information when designing DSRV systems.
\end{abstract}


\section{Introduction}
\label{s1}

The imitation of everyday sounds using the human voice is an efficient way of communicating sound concepts and ideas. For this reason, there is an increasing interest in the audio and speech community for applications that can perform automatic \textit{Query by Vocal Imitation} (QVI) \cite{lemaitre2016vocal} with the primary intention of (but not limited to) assisting search engines. In the music industry in particular, the vocal imitation of musical instruments often plays an important role in the production process, e.g. by allowing artists to tell producers how they like different instruments to sound in the mix \cite{mehrabi2017vocal}. This is especially true nowadays, with sound libraries of sampled musical instruments being used in a big proportion of music genres including popular music.

Commercial drum sound libraries typically contain thousands of samples from different drum types and articulations. These libraries could be too large for an individual to explore and find, for example, a specific snare drum sample to use in a musical piece. To save search time, some pieces of software use perceptual attributes like ``brightness'' or ``warmness'' to filter the samples \cite{zhang2019sound}. Drum Sample Retrieval by Vocalisation (DSRV) provides a complementary and potentially faster way of selecting the desired drum samples \cite{zhang2019sound, lemaitre2014effectiveness}, allowing artists to continue creating while avoiding exhaustive manual searches.

Research in QVI has been very active over the last decade and it is mainly characterised by investigations on (i) the nature of vocal imitations and the acoustic relationship with their respective source sounds \cite{lemaitre2016vocal, mehrabi2019vocal, delgado2020spectral}, (ii) the ability of humans to imitate sounds and link imitations to their source sound \cite{cartwright2015vocalsketch, lemaitre2016vocal, mehrabi2017vocal}, (iii) the usefulness of vocal imitations for sound design \cite{lemaitre2013non, cartwright2014synthassist}, (iv) the suitability of vocal imitations as inputs for audio retrieval engines \cite{zhang2018siamese, zhang2020vroom}, and (v) the relative importance of vocal imitations when compared to text information \cite{lemaitre2014effectiveness, zhang2020vroom}. Insights from these studies suggest that humans are generally skilled when performing and recognising vocal imitations of generic sounds, with most results pointing towards the high retrieval accuracy, speed, and usability of QVI systems when compared to query-by-text ones. Recent implementations of QVI systems use \textit{deep representation learning} techniques to estimate the similarity between source sounds and vocal imitations through metric learning, which significantly increases the retrieval accuracy of QVI systems \cite{zhang2018siamese, zhang2020vroom, zhang2016imisound, mehrabi2018similarity}.

Drum Sample Retrieval by Vocalisation (DSRV) has been implicitly studied in the past through QVI, as drum samples were often part of the evaluation datasets. DSRV was firstly studied in \cite{mehrabi2018similarity}, where authors implemented a convolutional autoencoder model to extract embeddings (learnt feature sets) that can predict human listeners' drum-imitation similarity scores. They later inspected their data and saw that it was difficult for some listeners to reproduce their own similarity scores for certain drum-imitation pairs and that imitators tended to focus on the sounds' envelope when imitating same-instrument sounds \cite{mehrabi2019vocal}. This observation was confirmed in \cite{delgado2020spectral}, with both studies remarking the potential usefulness of user-driven algorithms.





This study can be seen as a continuation of the above-mentioned work featuring convolutional autoencoders \cite{mehrabi2018similarity}. As our work is specifically directed to improve DSRV applications, we included two acoustics-based drum-imitation similarity metrics to complement the two perception-based similarity metrics in Mehrabi et al.'s paper. Acoustics-based metrics differ from perception-based ones in that the former investigate the correspondence of variations in the acoustic space of drum sounds with those in the acoustic space of vocal imitations, while the latter investigate the correspondence of drum-imitation feature distances with human listeners' drum-imitation similarity ratings. Apart from the inclusion of such metrics, we (i) explored three types of label conditioning, (ii) investigated how networks' architectures and hyperparameters affected performance, (iii) explored imitators' differences in vocal imitation styles, (iv) included new drum and vocal percussion datasets for training purposes, and (v) modelled single hits from vocal percussion and drum sounds exclusively, in particular those related to kick drum, snare drum, closed hi-hat, and opened hi-hat.


In Section \ref{s2}, we present the dataset of drum sounds and vocal percussion used throughout the study and detail the implementation of the proposed deep variational models, baseline methods, and evaluation metrics and routines. We present the final results and discuss their implications for DSRV in Section \ref{s3}, and outline the final conclusions of the study in Section \ref{s4}.

\section{Methodology}
\label{s2}

The main goal of this study was to discover audio embeddings that can best link vocal imitations with their reference drum samples both acoustically and perceptually. This section details the methodology followed in the present work to get the above-mentioned embeddings, including routines for data curation and pre-processing, model design and training, and final evaluation. The implementation of these can be found in our project's repository\footnote{https://github.com/alejandrodl/drum-sample-retrieval-vocalisation}.


\subsection{Data and Pre-Processing}
\label{s21}

We used several open-source and commercial datasets of drum samples and vocal percussion performances to train and evaluate our proposed algorithms. These are gathered in Table \ref{t1}. All drum samples and vocal percussion sounds had associated a label relative to the drum instrument they described or emulated: kick drum, snare drum, closed hi-hat, and opened hi-hat.

Training files relative to \textit{vocal imitations} (9,410 sounds) were taken from three vocal percussion datasets: \textit{AVP} \cite{delgado2019new}, \textit{BS1} \cite{stowell2010delayed}, \textit{BTX} \cite{zhu2020study}, and \textit{LVT} \cite{ramires2018user} (second and third subsets). In the case of the AVP dataset, we used the sounds contained in the personal subset and in seven improvisation performances from the fixed dataset, specifically those from participants 4, 8, 12, 16, 20, 24, and 28. We also exclusively used the samples in the BS1 dataset that were annotated as kick drums, snare drums, closed hi-hats, and opened hi-hats. The \textit{real drum samples} (8,876 sounds) were taken from miscellaneous sound libraries containing both acoustic and electronic sources. We only took the files whose names or directories reveal their class label (e.g. ``snare01.wav'', ``SN01.wav'', ``/snares/01.wav''...).


The files we used to evaluate the embeddings were taken from the \textit{VIPS} dataset \cite{mehrabi2018similarity}. This included 30 drum sounds from acoustic and electronic sources and their respective vocal imitations performed by 14 participants. After removing the samples relative to cymbals and toms, we had 18 drum samples and 252 vocal imitations (18 per participant). Reference sounds included 6 kick drums, 6 snare drums, 2 closed hi-hats, and 4 opened hi-hats. Silences sections were trimmed for all sounds.

\begin{table}
\centering
\setlength\tabcolsep{6pt}
\begin{tabular}{|c|c|c|c|} \hline
\textbf{Dataset}    & \textbf{Type} & \textbf{Usage} & \textbf{\# Samples} \\ \hline
Misc      & Drums    & Training       & 7,900        \\ \hline
BFD3 \cite{BFD}        & Drums    & Training       & 976         \\ \hline
AVP \cite{delgado2019new}         & Imitations    & Training       & 5,220         \\ \hline
BS1 \cite{stowell2010delayed} & Imitations    & Training       & 2,317         \\ \hline
LVT \cite{ramires2018user}         & Imitations    & Training       & 1,682        \\ \hline
BTX \cite{zhu2020study} & Imitations    & Training       & 191         \\ \hline
VIPS \cite{mehrabi2018similarity}        & Both          & Evaluation     & 270 \\ \hline
\end{tabular}
\caption{\label{t1}Datasets of drum sounds and vocal imitations used throughout this study.}
\end{table}


We applied an 8-fold \textit{waveform data augmentation} \cite{nanni2020data} to drum samples and vocal percussion sound events, specifically random \textit{pitch-shifting} (semitone range = [-1.0,+1.0] for drums and [-1.5,+1.5] for vocal imitations) and \textit{time-stretching} (stretch factor range = [0.7,1.3] for drums and [0.8,1.2] for vocal imitations), one after the other in random order. After this, we calculated the Bark spectrogram using a frame size of 46 ms and a hop size of 11 ms and applied the Terhardt’s ear model curves \cite{terhardt1979calculating} to scale the spectrograms to decibels. Spectrograms had dimensions of 128 frequency bins by 128 frames ($\sim$ 1.5 s).



\subsection{Baseline Methods}
\label{s22}

We built a \textit{heuristic feature set} made from the first 12 Mel Frequency Cepstral Coefficients (MFCC) excluding the zeroth coefficient, their first derivatives, the duration, the derivative after the envelope's maximum (DerAM), and the mean and standard deviation of the loudness and the pitch, calculated using the time-domain YIN method \cite{de2002yin}. The MFCCs, which are widely used in timbre analysis \cite{de2012enhancing, terasawa2005thirteen, terasawa2005perceptual}, were computed using 40 Mel bands, a frame size of 46 ms, and a hop size of 11 ms.


We also implemented the regular \textit{Convolutional Autoencoder} (CAE) model in \cite{mehrabi2018similarity}. We reproduced the architecture and the training routine of the best-performing model as reference. As we are evaluating sets of 32 features, we added a final fully-connected layer to the original model that connects its latent space of size 128 to an adapted one of size 32.

\subsection{Proposed Models}
\label{s23}
 
We adopted the CAE in \cite{mehrabi2018similarity} as the base architecture of our proposed model. From there, we applied several modifications that we later evaluated in terms of performance both individually and jointly (see section \ref{s3}). These modifications to the original model included a variational latent space \cite{kingma2013auto}, a change in the number of filters per layer, the replacement of upsampling layers and convolutional layers with one unique transposed convolution in the decoder, the use of max pooling operations instead of strided convolutions in the encoder, and a change in the kernel dimensions of the filters. We explored these modifications ensuring that the networks had a similar amount of parameters.


Once our final proposed model was built, we investigated the impact of the inclusion of label information in the model's architecture. This is known as \textit{model conditioning} \cite{choi2019variable} and it has been proven beneficial in some audio feature learning tasks \cite{dahmani2019conditional, chettri2020deep}. We built three models with the same architecture as our proposed CAE but conditioned on different label sets. These sets were (i) \textit{Sound-type Labels} (SL) to distinguish between drums and vocal imitation (two labels); (ii) \textit{Drum-type Labels} (DL) to distinguish between kick drum, snare drum, closed hi-hat, and opened hi-hat (four labels); and (iii) \textit{Sound and Drum-type Labels} (SDL) to distinguish between kick drum, snare drum, closed hi-hat, and opened hi-hat and whether they are drums or vocal imitations (eight labels). We fed the labels by concatenating a one-hot encoded vector to the flattened feature maps after the last convolutional block in the encoder and to the ones before the first convolutional block in the decoder.


Models were trained using an Adam optimisation algorithm \cite{kingma2014adam}, early stopping if validation loss has not decreased after 10 epochs, and downscaling of the learning rate by a factor of 0.2 if validation loss has not decreased after 5 epochs.

\subsection{Similarity Metrics}
\label{s24}

An ideal DSRV algorithm would rank drum samples in the correct order of similarity to an input vocal imitation. Measuring this similarity, however, is far from direct. If imitations are associated with a single sound, like in our case, one can use QVI metrics like the mean reciprocal rank to evaluate systems \cite{zhang2015retrieving, zhang2016imisound, zhang2018siamese}, although it does not assess the similarity of lower-ranked samples with the imitation. To account for the latter, other metrics and annotations like drum-imitation perceptual similarity ratings from listeners are indispensable \cite{mehrabi2018similarity}.

With this in mind, we evaluated the relevance of our feature sets via four similarity metrics. On the one hand, the \textit{Mean Reciprocal Rank} (MRR) \cite{muller2015fundamentals} and the \textit{Mantel Score Significance} (MSS) \cite{smouse1986multiple} measure similarity between sounds and vocal imitations by correspondances between both acoustic spaces. We refer to these as \textit{acoustics-based} similarity metrics. On the other hand, the \textit{Akaike Information Criterion} (AIC) \cite{bozdogan1987model} and the prediction \textit{accuracy} measure the capacity of feature sets to model and predict human listeners' drum-imitation similarity scores \cite{mehrabi2018similarity}. We refer to these as \textit{perception-based} similarity metrics.

\subsubsection{Acoustic-based Metrics}
\label{s241}

For a collection of $N$ vocal imitations per imitator (18 in our case), the Mean Reciprocal Rank (MRR) is defined as:
\begin{equation}
\label{eq:MRR}
\mathrm{MRR}=\frac{1}{N} \sum_{n=1}^{N} \frac{1}{\operatorname{rank}_{n}}
\end{equation}
, where $rank_{n}$ is the rank of the relevant reference drum sound for the $n$-th imitation. As the MRR is inversely proportional to the rank number, it acts as a penalisation factor in a simulated DSRV task: the more delayed the retrieval of the correct sample, the lower the MRR score.

The second acoustics-based metric, the Mantel Score Significance (MSS), measures the degree of global correspondence between the acoustic space of drum samples and that of vocal imitations for all imitators. To calculate it, we ran the Mantel test \cite{smouse1986multiple} on the individual features that compose embeddings. We constructed two Euclidean distance matrices for each individual feature and each individual imitator given by:
\begin{equation}
D_{ij}^{\mathit{ref}}=\left |x_{i}^{\mathit{ref}}-x_{j}^{\mathit{ref}}\right|^{2}
\end{equation}
\begin{equation}
D_{i j}^{\mathit{imi}}=\sum_{n=1}^{N}\left|x_{n,i}^{\mathit{imi}}-x_{n,j}^{\mathit{imi}}\right|^{2}
\end{equation}
, where $i\in[1,18]$ and $j\in[1,18]$ are the indices of the drum sound classes, $N$ is the number of imitators (14 in our case), ${x}^{\mathit{ref}}$ are the feature magnitudes from the reference drum sounds, and ${x}^{\mathit{imi}}_{n}$ are the feature magnitudes from the vocal imitations pertaining to the $n$-th user with $n\in[1,14]$. We ran the Mantel test on these two matrices, measuring the degree of statistical (Pearson) correlation between them with the associated p-value. We reported the percentage of performances whose Mantel test scores were statistically significant (p$<$0.05). We ran 5 Mantel tests per feature and later averaged these results.


\subsubsection{Perception-based Metrics}
\label{s242}

We followed the same procedure as in Mehrabi et al. \cite{mehrabi2018similarity} to calculate the two remaining metrics. In this study, 63 human listeners rated the perceived similarity between a vocal imitation and same-category drum sounds within the VIPS dataset. They were presented with 30 test pages (trials), of which 28 were unique and 2 were random duplicates used to assess listeners' reliability. The listeners that were able to replicate their responses for at least one of the duplicates with a Spearman rank correlation coefficient of $\rho\geq0.5$ were considered reliable. In this study, we got 51 reliable listeners and a total of 5,532 similarity ratings. For each feature set, the Euclidean distance was measured between each of the 252 imitations and their respective 6 within-class sounds, giving 1512 distance values. We calculated the percentage of imitated sounds for which the fitted slope of the regression model $rating \sim distance$ was significantly below 0. This is an implicit measure of \textit{accuracy}, with a negative slope meaning an inverse relationship between distances and similarity ratings. A slope's value is significantly below 0 if the upper 95\% confidence interval (CI) is negative.


One of the limitations of the regression model above mentioned is that it only takes sounds' features and similarity measures into account. It, therefore, fails to capture the influence of other determinant factors like which imitator produced the current sound or whose listener is the current similarity score. To account for these idiosyncrasies, we fitted a Linear Mixed-Effect Regression (LMER) model using R's lme4 package \cite{bates2014fitting} which was specified as $rating \sim distance * sound + (1|listener/trial) + (1|imitator)$, with the similarity \textit{rating} as the response variable, fixed effects of \textit{distance} and drum \textit{sound} with an interaction term, and random intercepts for \textit{listener}, \textit{imitator}, and \textit{trial} (nested in listener). We used the \textit{Akaike Information Criterion} (AIC) as a metric to evaluate model fitness. Given two models, the one with the lowest AIC displays less information loss and is, thus, a better fit. While the absolute value of the AIC has no statistical meaning, a practical heuristic to compare scores is to consider two models with $\Delta_{AIC}>10$ as one fitting the data significantly better than the other \cite{anderson2004model}.

Finally, note that results for these two perception-based metrics are not directly comparable to those in \cite{mehrabi2018similarity}, as we did not analyse toms and cymbals, used a lower embedding size, used a different dataset containing instrument labels, and averaged results over five models per method.




\subsection{User Differences in Vocal Imitation Style}
\label{s25}

To conclude the study, we briefly investigated differences in vocal imitation styles among imitators. For this, we followed the methodology outlined in \cite{delgado2020spectral}, which consisted in performing the Mantel test as described in Section \ref{s242} and plotting a heatmap containing the p-values of these tests. The lower the p-value a feature has for a certain imitator, the more faithfully the imitator reproduces such feature vocally. We study the heuristic feature set, which contains acoustically interpretable features, and compare its results with those from the best-performing feature set.

\section{Results and Discussion}
\label{s3}

Applying the modifications in \ref{s23} to the baseline model (CAE-B) \cite{mehrabi2018similarity}, we found that the model benefited from (i) an increase in the number of filters from 8-16-24-32 to 8-16-32-64, (ii) a change in internal filter sizes from 10x10 to 9x9, which avoids aliasing from odd dimensions, and (iii) max-pooling operations in the encoder. The use of a variational latent space, an increased depth of the networks, and the use of transposed convolutions instead of upsampling layers in the decoder had a minor effect on performance while sometimes lowering the quality of sample reconstructions. Hence, we kept the regular latent space and upsampling layers in our proposed model (CAE).


\begin{table}
\centering
\setlength\tabcolsep{2.2pt}
\renewcommand{\arraystretch}{1}%
\begin{tabular}{cccccccc}
& \multicolumn{2}{c}{Baseline (\ref{s22})} & \multicolumn{4}{c}{Proposed (\ref{s23})}\\
\cmidrule(l){2-3} \cmidrule(l){4-7}
& \mc{HS} & \mc{CAE-B}  & \mc{CAE}  & \mc{CAE-SL} & \mc{CAE-DL} & \mc{CAE-SDL}\\
\midrule
MRR   & 34.7      & 39.6$|$1.2     & \textbf{42.1$|$1.0}          & 41.2$|$1.0          & 41.4$|$1.3          & 41.7$|$0.6 \\
\addlinespace
MSS   & 27.6      & 44.2$|$3.3      & 43.0$|$3.1           & 44.8$|$3.0           & 45.0$|$2.2          & \textbf{47.5$|$2.5} \\
\midrule
Acc.     & \textbf{55.6}      & 43.3$|$1.9       & 47.8$|$3.9         & 48.9$|$5.7           & 47.8$|$5.8          & 54.4$|$1.9 \\
\addlinespace
AIC      & 999       & 658$|$51        & 625$|$53           & 609$|$56            & 613$|$51           & \textbf{576$|$58} \\
\midrule
\end{tabular}
\caption{\label{t2} Results for the four evaluation metrics. Except for the heuristic set's results (HS), two numbers separated by a vertical bar are reported: the mean scores (left) and 95\% confidence intervals (right) from the five different models' performances. MRR is expressed in terms of percentage and AIC values are substracted a constant value to improve readability. Numbers in bold indicate best performances for the related metric.}
\end{table}

Final results for all evaluation metrics are displayed in Table \ref{t2}. We see that the four proposed CAE models' results equal or surpass the ones from the baseline CAE-B model. Also, conditional models tend to perform better than their non-conditional counterpart (CAE) and the CAE-SDL model performs better than or similar to the rest of methods in all metrics.




Regarding \textit{acoustics-based} similarity metrics, MRR and MSS, CAE embeddings performed consistently better than the heuristic feature set. Similar results were observed in \cite{zhang2015retrieving} for MRR, where learnt features from a stacked autoencoder performed better than a set of 13 MFCCs and their first and second derivatives. We also noted relatively high MSS scores for CAE embeddings, reaching 47.5\% in the case of CAE-SDL. This means that statistically speaking, a generic imitator would dexterously reproduce with the voice 15 out of the 32 CAE-SDL features as heard in the reference drum sounds.




Regarding \textit{perception-based} similarity metrics, accuracy and AIC, we find that deep embeddings from CAE models perform significantly better than heuristic features in AIC but the opposite is true in the case of the accuracy metric. This is a rather surprising result considering the higher accuracy scores by learnt features in \cite{mehrabi2018similarity}, which could be due to the differences in experimental setting between the study and ours (see section \ref{s24}). AIC scores from heuristic features are far from CAE-related ones, as also observed in \cite{mehrabi2018similarity}, which might mean that heuristic features do not benefit as much as CAE embeddings from knowing about the current listener, imitator, drum sound, and trial variables when computing similarity. This could be due to the fact that CAE embeddings are learnt using datasets containing a wide range of vocal percussion and drum sounds from various performers while heuristic features are extracted from sounds in a context-agnostic way. We also see that the CAE-SDL accuracy score (54.4\%) is relatively close to that of the heuristic set (55.6\%). The CAE-SDL's AIC score is also the lowest one, although 95\% confidence intervals from AIC scores highly overlap with each other, meaning that we cannot ensure the superiority of one method over the rest. Increasing the number of models per method might clear doubts in this respect.

Apart from results in Table \ref{t2}, we also trained the CAE-B model with the dataset used in the original study \cite{mehrabi2018similarity}, which gave scores of 39.2$|$1.2, 45.8$|$3.7, 48.9$|$6.5, and 579$|$94 for MRR, MSS, accuracy, and AIC respectively. It, therefore, outperformed the CAE-B trained with our dataset in perception-based metrics but not the best-performing model (CAE-SDL) in any metric. Also, all methods significantly outperformed random feature vectors (19.3$|$0.9, 4.7$|$1.6, 7.8$|$3.9, and 1035$|$129).

In summary, results show that conditioning DSRV systems on sound- and drum-type labels is very likely to augment the degree of similarity between imitations and retrieved samples. We believe these results can be extrapolated to general query by vocal imitation to some degree and that QVI algorithms can potentially benefit from this kind of conditioning, including metric learning systems trained with sample-imitation pairs \cite{zhang2018siamese}.

\begin{figure}[t]
\begin{center}
\includegraphics[width=1\columnwidth]{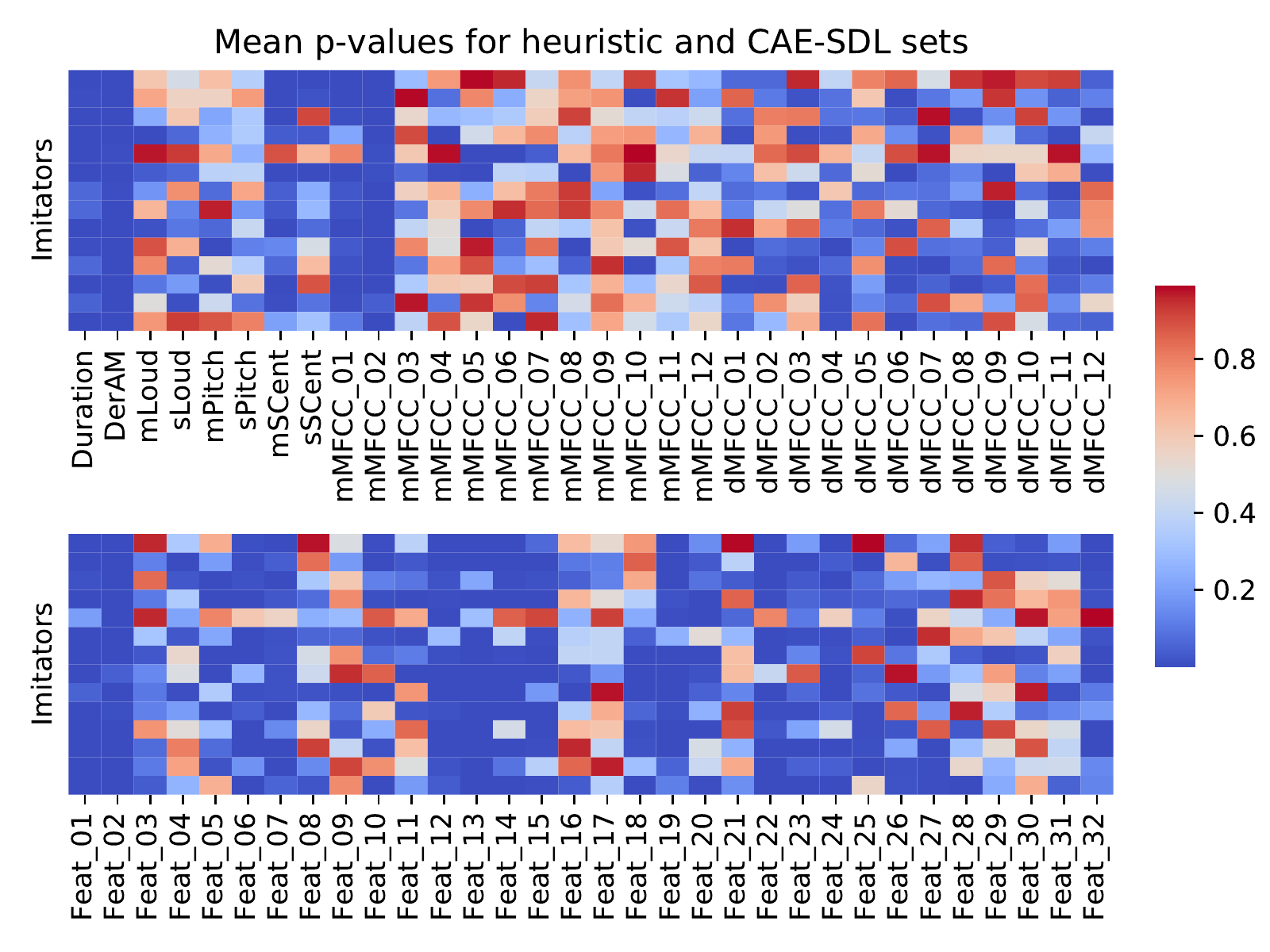}
\caption{Mean Mantel test p-values (5 iterations) for each imitator using the features in the heuristic set (above) and one of the CAE-SDL embeddings (below). DerAM = envelope's derivative after maximum; Loud = loudness; SCent = spectral centroid; m = mean; s = standard deviation; d = derivative.}
\label{f1}
\end{center}
\end{figure}


Finally, we explored user differences through Figure \ref{f1}, which illustrates the mean p-values from the five iterations of the Mantel test using the heuristic feature set and one of the five CAE-SDL embeddings. We can see how the p-values associated with CAE-SDL features are generally lower than those relative to the heuristic set, which is also reflected in their MSS scores in Table \ref{t2}. The p-values related to the derivative after the maximum of the envelope are very low for all imitators, lying in accordance with observations in \cite{delgado2020spectral}. It also relates to the finding in \cite{mehrabi2017vocal} that imitators tend to reproduce descending loudness ramps with high accuracy. We also note low p-values for the mean of the spectral centroid, MFCC 1, and MFCC 2. These relate to the energy ratio between low and high frequencies and the perceived ``brilliance" of the sounds \cite{caetano2019audio}. The low p-values for the duration of the sound indicate that this feature was also emulated dexterously by all imitators, which was found to be of critical importance in \cite{lemaitre2016vocal} for generic sounds. The same study and \cite{mehrabi2017vocal} found that pitch was a key feature in vocal imitations, although it does not appear to be the case for drum sounds. Finally, while some heuristic and learnt features are imitated skillfully by all imitators, many other features find notable disagreements between imitators across both heatmaps. This further supports the idea of approaching DSRV via user-based systems \cite{gong2020contextual}, which can fine-tune retrieval scores to specific imitators by taking their vocal imitation styles into account.

\section{Conclusion}
\label{s4}

We have explored the potential of several types of deep convolutional autoencoder models to learn useful feature sets for DSRV. We used four different interpretable metrics to investigate how well these embeddings could link vocal imitations with their reference drum sounds and found that models conditioned on both sound- and drum-type labels (CAE-SDL) excelled in both acoustics- and perception-based metrics. Future work will explore ways to fine-tune predictions by building user-based DSRV systems based on the CAE-SDL embeddings.

\section{Acknowledgements}

This work has received funding from the European Union’s Horizon 2020 research and innovation programme under the Marie Skłodowska-Curie grant agreement No. 765068.

\bibliographystyle{IEEEtran}

\bibliography{Interspeech2022}


\end{document}